\documentclass[]{JHEP3}
\usepackage{bm}
\usepackage[]{youngtab}
\usepackage{epsfig}
\usepackage{pstricks}

\def\nn{ \nonumber \\ }
\def\ket#1{ \left| #1 \right\rangle }
\def\text#1{\hbox{#1}}
\def\openone{\rlap{1}\hspace{1.5pt}1}
\def\agt{\gtrsim}
\def\vev#1{ \left\langle #1 \right\rangle }
\def\bra#1{ \left\langle #1 \right| }
\def\abs#1{ \left| #1 \right| }

\def\xslash#1{ \rlap{/}{#1} }

\title{ 1/N Expansion for Exotic Baryons}

\author{Elizabeth~Jenkins and Aneesh V. Manohar\\ 
Department of Physics, University of California at San Diego,  La Jolla, CA 92093, USA}

\abstract{The $1/N_c$ expansion for exotic baryons is developed, and applied to the masses, meson couplings and decay widths. Masses and widths of the $\mathbf{27}$ and $\mathbf{35}$ pentaquark states in the same tower as the $\mathbf{\overline{10}}\supset \Theta^+$ are related by spin-flavor symmetry.  The $\mathbf{27}$ and $\mathbf{35}$ states can decay within the pentaquark tower, as well as to normal baryons, and so have larger decay widths than the lightest pentaquark $\Theta$.  The $1/N_c$ expansion also is applied to baryon exotics containing a single heavy antiquark.  The decay widths of heavy pentaquarks via pion emission, and 
to normal baryons plus heavy $\overline D^{(*)},B^{(*)}$ mesons are studied, and relations following from large-$N_c$ spin-flavor symmetry and from heavy quark symmetry are derived.}

\keywords{qcd, nex, pmo}

\preprint{UCSD/PTH 04-02}

\begin{document}

\section{Introduction}

The recent discovery of  the $\Theta^+$ baryon with strangeness $S=+1$~\cite{Thetadiscovery} has led to renewed interest in hadron spectroscopy. The $\Theta^+$ contains an $\bar s$ quark since it has strangeness $+1$, and is thought to be a five-quark $uudd\bar s$ state. The $q^4 \bar q$ states which include the $\Theta^+$ form a spin-1/2 flavor $SU(3)$ antidecuplet. The states in the $\overline{\mathbf{10}}$ are shown in Fig.~\ref{fig:2}, using the tentative notation proposed by the Particle Data Group.
\FIGURE{
\psset{xunit=0.75cm}
\psset{yunit=0.6495cm}
\psset{runit=0.3cm}
\psset{dotsize=0.2cm}
\begin{pspicture}(-5,-3)(4,5.5)
\psline[linestyle=dashed]{->}(-3,0)(4,0)
\psline[linestyle=dashed]{->}(0,-3)(0,5)
\rput(0,5.5){$Y$}
\rput(4.5,0){$I_3$}
\psline(3,-2)(0,4)(-3,-2)(3,-2)
\psdots[dotstyle=o](3,-2)(1,-2)(-1,-2)(-3,-2)(2,0)(0,0)(-2,0)(-1,2)(1,2)(0,4)
\rput(-4,4){$\Theta$}
\rput(-4,2){$N_{\overline{10}}$}
\rput(-4,0){$\Sigma_{\overline{10}}$}
\rput(-4,-2){$\Phi$}
\end{pspicture}
\caption{ \label{fig:2} $SU(3)$ weight diagram for the $\overline{\mathbf{10}}$ baryons.}
}

The states at the corners of the triangle are manifestly exotic, since they do not have the quantum numbers of $qqq$ states. The observed $\Theta^+(1540)$~\cite{Thetadiscovery} is the isosinglet state at the top of the triangle. 
The $I = 3/2$ states $\Phi(1860)$~\cite{xiexotics} (usually called $\Xi_{3/2}$) also have been observed and are thought to be the $ssqq \bar q$ members of the antidecuplet.
The experimental situation is still confusing, with some experiments reporting discoveries, and others reporting upper limits on the production cross-section for the states. There is also an upper bound on the width of the $\Theta^+$ derived in Ref.~\cite{Cahn}. In this article, we derive results for exotic baryons which can be obtained in a systematic expansion in $1/N_c$, where $N_c$ is the number of colors. Readers  not interested in technical details of the calculation can skip to the phenomenological results, which start at Sec.~\ref{sec:masses}.
  
The $1/N_c$ expansion of QCD~\cite{'tHooft:1973jz} constrains the spin-flavor properties of baryons and their couplings to mesons.  In the $N_c \to \infty$ limit,  baryons have an exact contracted spin-flavor symmetry~\cite{Largenspinflavor}, which can be used to classify states.  This spin-flavor symmetry is broken at subleading orders in $1/N_c$.  The $1/N_c$ expansion is a systematic expansion for QCD, and gives  results  for the ground-state $[\mathbf{56},0^+]$ baryons ($\mathbf{8}_{1/2}$ and $\mathbf{10}_{3/2}$) in excellent agreement with experiment, with some predictions accurate to a fraction of a percent~\cite{Largenspinflavor,DJM1,DJM2,largenrefs,JenkinsLebed,largen}. The $1/N_c$ expansion also has been successful in explaining the spin-flavor properties of  heavy-quark baryons~\cite{heavybaryons} and excited baryons~\cite{seventyminusrefs}. 
In this paper, we extend the $1/N_c$ expansion to the lowest-lying exotic baryon states identified in Ref.~\cite{JM1} using the method of quark operators~\cite{DJM1,DJM2,largen}. We will concentrate on predictions in the isospin and $SU(3)$ flavor symmetry limits for the exotic baryon masses, axial couplings and decay widths.  A complete $SU(3)$ analysis including $SU(3)$-breaking will be presented in a longer publication~\cite{JM3}. Implications of large-$N_c$ for exotic baryons have also been considered in Ref.~\cite{CohenLebed}.

In the $N_c \to \infty$ limit, baryons form irreducible representations of contracted spin-flavor symmetry $SU(6)_c$, and all states in a given irreducible representation (tower) are degenerate in the $N_c \to \infty$ limit~\cite{DJM1}. 
For finite $N_c$, spin-flavor symmetry breaking generates a mass-splitting between the baryons in a given tower proportional to $\mathbf{J}^2/N_c$~\cite{Largenspinflavor}.
The irreducible representations have been constructed as induced representations~\cite{DJM1}, and are referred to as the Skyrme representations, since their group-theoretic structure is identical to that in the Skyrme model. For finite $N_c$, it is convenient to work with irreducible representations of uncontracted spin-flavor symmetry $SU(6)$, which are finite towers and are referred to as quark representations~\cite{DJM1,largen,DJM2}. The quark and Skyrme representations are completely equivalent for $N_c \to \infty$~\cite{Manohar:1984}. For finite $N_c$, the baryon tower of the Skyrme representation remains infinite and continues to satisfy the $SU(6)_c$ algebra. The quark representation gives a finite set of states for finite $N_c$, and satisfies the $SU(6)$ algebra. QCD baryons satisfy the $SU(6)_c$ algebra for $N_c \to \infty$; the $SU(6)$ algebra for finite $N_c$ is not satisfied in QCD. The difference between Skyrme and quark representations is a subleading $1/N_c$ effect which can be absorbed into redefinitions of the the unknown coefficients of $1/N_c$ suppressed operator products.  For finite values of $N_c$,  the two yield equivalent group theoretic results to any given order in $1/N_c$.

Large-$N_c$ spin-flavor symmetry does not predict that exotic baryons exist.  However, if exotic baryons do exist, consistency of the $1/N_c$ expansion relates the spin-flavor properties of exotic and non-exotic baryons at order $1/N_c$, just as it relates the spin-flavor properties of heavy-quark and light-quark baryons~\cite{heavybaryons}.
These same conclusions also are obtained in the Quark and Skyrme Models, which share the same contracted spin-flavor symmetry in the large-$N_c$ limit.
In the Quark Model, exotic baryons with exoticness\footnote{Exoticness is defined as the minimal value of $E$ for which the baryon flavor representation can be constructed from $qqq(\bar q q)^E$. See Ref.~\cite{JM1,JM4} for a more extensive discussion, and  for the definition of exoticness in the Skyrme model, which does not have explicit quark degrees of freedom.} $E$ form distinct irreducible representations of $SU(6)$.  In the Skyrme model, exotics contribute at $1/N_c$~\cite{DJM2}, and cannot be separated from $1/N_c$ corrections.  In both cases, one cannot prove that exotics must exist as $N_c\to \infty$.

In this paper, our main focus is on the lowest-lying baryon exotics, the pentaquarks. The $E=1$ pentaquarks consist of the $\mathbf{\overline{10}}_{1/2}$, $\mathbf{27}_{1/2}$, $\mathbf{27}_{3/2}$, $\mathbf{35}_{3/2}$ and $\mathbf{35}_{5/2}$ representations of flavor and spin, and have positive parity. The results in this paper do not include $SU(3)$ breaking. Thus, results for masses and widths are average values for all the states in a given $SU(3)$ flavor representation. The $1/N_c$ expansion also can be performed using only isospin $SU(2)$ flavor symmetry, with no assumption that the strange quark mass is small. The results obtained using isospin symmetry are similar to the results for three flavors in the limit of $SU(3)$ symmetry.
The $SU(3)$ pentaquark towers decompose into isospin towers with fixed strangeness in the $SU(2)$ flavor symmetry limit.  The $S=1$ states are the $\Theta^+$ in the $\mathbf{\overline{10}}$, the isotriplets $\Theta_{1,j=1/2}$ and $\Theta_{1,j=3/2}$ in the $\mathbf{27}_{1/2}$ and $\mathbf{27}_{3/2}$, and the isotensors $\Theta_{2,j=3/2}$ and $\Theta_{2,j=5/2}$ in the $\mathbf{35}_{3/2}$ and 
$\mathbf{35}_{5/2}$. The individual isotriplet states are the $\Theta_1^{++}$, $\Theta_1^{+}$ and $\Theta_1^0$, and the individual isotensor states are $\Theta_2^{+++}$, $\Theta_2^{++}$, $\Theta_2^{+}$, $\Theta_2^{0}$ and $\Theta_2^-$.  Unlike $SU(3)$ flavor-symmetry breaking, isospin flavor-symmetry breaking can be neglected in comparison to $1/N_c$ corrections. 

\section{The $1/N_c$ expansion for exotics} 

Baryons of exoticness $E$ have $N_c+E$ quarks in the completely symmetric $SU(6)_q$ representation, and $E$ antiquarks in the completely symmetric $SU(6)_{\bar q}$ representation~\cite{JM1}. In a quark model description, $E$ of the $N_c+E$ quarks must be in $\ell=1$ orbitally excited wavefunctions, so that the total wavefunction satisfies Fermi statistics by being completely antisymmetric in color-orbital space~\cite{Glozman}.  Including the orbital angular momentum, the excited quarks $q^*$ can be either $j=1/2$ or $j=3/2$.  Whether the quark spin is actually made up of quark spin plus orbital angular momentum, or just plain quark spin, is irrelevant.  All that matters for the spin-flavor algebra is the spin-flavor structure of the state. The irreducible representation we choose is the one with $q^*(j=1/2)$ in a completely symmetric spin-flavor state with the other unexcited quarks.  These states have positive parity. The other symmetry structures correspond to choosing different irreducible representations of $SU(6)_c$. The mixed symmetry spin-flavor representation with all quarks in the ground state $\ell=0$ wavefunction have negative parity~\cite{Pirjol}. Whether these are lighter is a dynamical issue. It has been argued that these are the lightest states for heavy pentaquarks~\cite{sww}.

The large-$N_c$ spin-flavor algebra of baryon exotics can be written in terms of one-body quark and antiquark operators.  Define bosonic quark and antiquark creation and annihilation operators $q^\dagger_{\cal A}$, $q^{\cal A}$, $\bar q^{\dagger {\cal A}}$ and $\bar q_{\cal A}$ with spin-flavor index ${\cal A}=1,\ldots,2F$ which satisfy the bosonic commutation relations
$\left[ q^{\cal A},q^\dagger_{\cal B} \right] = \delta^{\cal A}_{\cal B}$ and $\left[ \bar q_{\cal A}, \bar q^{\dagger {\cal B}} \right] = \delta^{\cal B}_{\cal A}$.  The $q$ and $\bar q$ creation and annihilation operators commute with one another.
Note that the creation/annihilation operators of the quark representation do not carry color indices. Baryon exotics are color singlets, so it is not necessary to keep track of color for the spin-flavor analysis. The $SU(2F)$ spin-flavor matrices in the $\yng(1)$ and $\overline{\yng(1)}$ representation are denoted by $\Lambda^A$ and $\bar \Lambda^A=-(\Lambda^A)^T$, respectively, with normalization $\text{Tr}\, \Lambda^A \Lambda^B=\delta^{AB}/2$.  Under the decomposition $SU(2F) \supset SU(F) \otimes SU(2)$ of the spin-flavor group to its flavor and spin subgroup, the $SU(2F)$ fundamental index ${\cal A} \to {\imath\alpha}$, where  $\imath=\uparrow\downarrow$ is a spin index and $\alpha=1,\ldots,F$ is a flavor index. The $SU(F)$ generators in the $\yng(1)$ and $\overline{\yng(1)}$ representations are $\left(T^a\right)^\alpha{}_\beta $ and $\left(\bar T^a\right)_\alpha{}^\beta$, respectively, with $ \bar T^a = -(T^a)^T$ and $\text{Tr}\, T^a T^b=\delta^{ab}/2$, and the spin generators are $\left(J^i\right)^\imath{}_\jmath=\left(\sigma^i\right)^\imath{}_\jmath/2$.  [For $SU(2)$, the $\mathbf{\bar 2}$ is equivalent to the $\mathbf{2}$, so it is possible to also use $J^i$ for the antiquark spin generators.]

The one-body quark and antiquark operators $\Lambda_q$ and $\Lambda_{\bar q}$ are defined by
\begin{eqnarray}
\begin{array}{rcl}
N_q &=& q^\dagger q,\\
J^i_q &=& q^\dagger \left( J^i \otimes \openone \right) q,\\
T^a_q &=& q^\dagger \left( \openone \otimes T^a \right) q,\\
G^{ia}_q &=& q^\dagger \left( J^i \otimes T^a \right) q,\\
\end{array}\qquad\qquad
\begin{array}{rcl}
N_{\bar q} &=& \bar q^\dagger \bar q,\\
J^i_{\bar q} &=& \bar q^\dagger \left( J^i \otimes \openone \right) \bar q,\\
T^a_{\bar q} &=& \bar q^\dagger \left( \openone \otimes \bar T^a \right) \bar q,\\
G^{ia}_{\bar q} &=& \bar q^\dagger \left( J^i \otimes \bar T^a \right) \bar q.\\
\end{array}
\label{2}
\end{eqnarray}
These one-body operators generate a $U(6)_q \times U(6)_{\bar q}$ spin-flavor algebra.  $1/N_c$ effects break the separate quark and antiquark spin-flavor symmetries down to the diagonal $SU(3) \times SU(2)$ subgroup of $SU(6)$.
The properly normalized diagonal $SU(6)$ generators are $(N_q-N_{\bar q})/(2 \sqrt F)$, $(J^i_q + J^i_{\bar q})/\sqrt F$, $(T^a_q + T^a_{\bar q})/\sqrt 2$ and $\sqrt 2(G^{ia}_q-G^{ia}_{\bar q})$.  Baryons with exoticness $E$ satisfy the identities $N_q=N_c+E$ and $N_{\bar q}=E$.

The Hamiltonian (or other baryon observable) can be written as an expansion~\cite{DJM1,largen,DJM2}
\begin{eqnarray}
H &=& \sum_n { {\cal O}_n \over N^{n-1}_c }
\label{3}
\end{eqnarray}
where the summation is over all independent $n$-body operators ${\cal O}_n$. 
For baryon exotics, the independent $n$-body operators are monomials of order $n$ in the basic one-body operators of Eq.~(\ref{2}). Operator reduction rules~\cite{DJM2} eliminate many operator products, considerably simplifying the $1/N_c$ expansion. For baryon exotics with quarks and antiquarks separately in completely symmetric spin-flavor representations, the quark operators satisfy the same identities given in Ref.~\cite{DJM2}, with the replacement $N_c \to N_q = N_c + E$, and the antiquark operators separately satisfy the same identities with $N_q \to - N_{\bar q}$, $J^i_q \to J^i_{\bar q}$, $T^a_q \to T^a_{\bar q}$ and $G^{ia}_q \to - G^{ia}_{\bar q}$, and
$N_c \to N_{\bar q} = E$.

Transition operators between baryons with different exoticness also are needed. The $\Delta E = -1$ one-body operators $\Lambda^A_-$ which annihilate one $q \bar q$ pair are
\begin{eqnarray}
\begin{array}{rcl}
N_- &=& \bar q q,\\
J^i_- &=& \bar q \left( J^i \otimes \openone \right) q,\\
\end{array}\qquad\qquad
\begin{array}{rcl}
T^a_- &=& \bar q \left( \openone \otimes T^a \right) q,\\
G^{ia}_- &=& \bar q \left( J^i \otimes T^a \right) q.\\
\end{array} 
\end{eqnarray}
The hermitian conjugate $\Delta E=1$ operators are $\Lambda_+^A = \Lambda_-^{A\dagger}$. 
The requirement that flavor-singlet quark-antiquark annihilation is forbidden for baryon exotics~\cite{JM1} implies that
\begin{eqnarray}
q^{\imath \alpha} \bar q_{\jmath \alpha} \ket{E}=0,\qquad \bra{E} q^\dagger_{\imath \alpha}\bar q^{\dagger \jmath \alpha}=0
\label{4}
\end{eqnarray}
where $\ket{E}$ denotes a baryon state with exoticness $E$. States which satisfy Eq.~(\ref{4}) will be called minimal states, and we will restrict our analysis to these states. Non-minimal states can be included by treating them as minimal states plus a $\bar q q$ pair~\cite{bosonic}. Eq.~(\ref{4}) implies that for minimal states
 \begin{eqnarray}
N_- =0,\quad J^i_- =0,\quad N_+ =0,\quad J^i_+ =0.
\label{11}
\end{eqnarray}
Operator products of $\Lambda_-$ and $\Lambda_+$ do not need to be considered, since these operators can be rewritten as linear combinations of $\Lambda_q$ and $\Lambda_{\bar q}$. Operators in a given $\Delta E $ sector are products of the form $\left(\Lambda_\pm\right)^{\abs{\Delta E}}$ times monomials in $\Lambda_q$ and $\Lambda_{\bar q}$.

There are new operator identities which reduce the number of independent operators products of the form $\Lambda_q \Lambda_{\bar q}$, $\Lambda_q \Lambda_-$, $\Lambda_{\bar q} \Lambda_-$ and $\Lambda_-\Lambda_-$, as well as the hermitian conjugate products involving $\Lambda_+$~\cite{JM3}.  The identities for an $SU(2)$ analysis correspond to a simple operator reduction rule: \emph{Operator products in which two flavor indices are contracted using $\delta^{ab}$ or $\epsilon^{abc}$ can be eliminated.}

In the $SU(3)$ limit, the most general $\Delta E=0$ Hamiltonian is a polynomial in the one-body operators divided by $N_c$. The operator reduction rules imply that $H$ has the form
\begin{eqnarray}
H_0 = N_c f\left( {J_q^2 \over N_c^2}, {J_{\bar q}^2 \over N_c^2}, {J^2 \over N_c^2}, {E \over N_c}\right),
\label{genform}
\end{eqnarray}
where the function $f$ is a general polynomial of its arguments.
The Hamiltonian to order $1/N_c$ is
\begin{eqnarray}
H_0  =  c_0 N_c + c_1 E + {1 \over N_c} \left( c_2 J_q^2  + c_3 J_{\bar q}^2 + c_4 J^2 +c_5 E^2\right) 
\label{20}
\end{eqnarray}
where the coefficients $c_i$  are functions of $1/N_c$ of order $N_c^0$, and exoticness $E$ is assumed to be order one. Although the Hamiltonian is symmetric under $q \leftrightarrow \bar q$, this symmetry is no longer manifest in Eq.~(\ref{20}) because the substituted values $N_q$ and $N_{\bar q}$ are asymmetric.  For example,
\begin{eqnarray}
{1\over N_c^2} \left[ N_q J_q^2 + N_{\bar q} J_{\bar q}^2 \right]
\end{eqnarray}
which is symmetric under $q \leftrightarrow \bar q$ becomes
\begin{eqnarray}
{1\over N_c^2} \left[(N_c+E) J_q^2 + E J_{\bar q}^2 \right] \to {1\over N_c} J_q^2
\end{eqnarray}
to order $1/N_c$ (assuming $E$ is order one). As a result, $c_2$ and $c_3$ in Eq.~(\ref{20}) need not be equal. 

The Hamiltonian in the Skyrme Model is given by Eq.~(\ref{20})
with the additional constraints~\cite{JM1}
\begin{eqnarray}
c_2=c_3=2c_5=2c_1,
\label{10}
\end{eqnarray}
if collective coordinate quantization with neglect of vibrational-rotational coupling is used.
Eq.~(\ref{10}) does not follow from a general $1/N_c$ analysis. The computation of the $\Theta^+$ mass in Ref.~\cite{DPP} used this restricted form for the Hamiltonian, but it has been shown that the computation in Ref.~\cite{DPP} is not a consistent semiclassical computation~\cite{IKOR,pobylitsa,cohen}. Including vibrational-rotational couplings gives a Hamiltonian which does not satisfy Eq.~(\ref{10}). The results in this paper use the general form of the $1/N_c$ Hamiltonian, and do not rely on Eq.~(\ref{10}).
 
 In the $SU(3)$ flavor symmetry limit, the most general $\Delta E \not=0$ Hamiltonian vanishes, and there is no mixing between states with different $\Delta E$.
To first order in $SU(3)$ breaking, the Hamiltonian is given by constructing all possible independent spin-singlet terms which transform as $T^8$. The $\Delta E = 0$ Hamiltonian is
\begin{eqnarray}
H^8 &=& c_1^{(8)} T^8_q + c_2^{(8)} T^8_{\bar q} + {E \over N_c}( c_3^{(8)} T^8_q + c_4^{(8)} T^8_{\bar q} ) \nn
&& + {1\over N_c} \biggl[ c_5^{(8)} J^i_q G^{i8}_q + c_6^{(8)} J^i_q G^{i8}_{\bar q} + c_7^{(8)} J^i_{\bar q} G^{i8}_{\bar q} + c_8^{(8)} J^i_{\bar q}  G^{i8}_{\bar q}\biggr]
\end{eqnarray}
to order $1/N_c$. This form of the $1/N_c$ expansion requires operator identities for the $SU(3)$ analyis, which are deferred to a future publication. Similarly, the $\Delta E=-1$ Hamiltonian is
\begin{eqnarray}
H^8_- &=& d_1^{(8)} T^8_- +  {1 \over N_c} \left[
d_2^{(8)} E T^8_- + d_3^{(8)} J^i_q G^{i8}_- \right]\ .
\label{40}
\end{eqnarray}
The $\Delta E = -1$ Hamiltonian mixes the pentaquark states  with the ordinary $E=0$ baryon states at order $m_s/\sqrt{N_c}$.  The $1/\sqrt{N_c}$ suppression factor is obtained because the matrix elements of $\Lambda_-$ operators are at most of order $\sqrt{N_c}$.

The $\Delta E=0$ axial currents are given by
\begin{eqnarray}
\frac 1 2 A^{ia}  &=& g_A G_q^{ia} +  g_1 G_{\bar q}^{ia}  + g_2 {{\{E, G_q^{ia}\}} \over N_c} +  g_3 {{\{E, G_{\bar q}^{ia}\}} \over N_c}\nn
&&   +  g_4 {J_q^i T^a_q \over N_c} +  g_5 {J_{\bar q}^i T^a_q \over N_c} + g_6 {J_q^i T^a_{\bar q} \over N_c} +  g_7 {J_{\bar q}^i T^a_{\bar q} \over N_c} 
\label{30}
\end{eqnarray}
up to operator products of relative order $1/N_c^2$. The leading term $G^{ia}_q$ is order ${N_c}$ for $a=1,2,3$, order $\sqrt{N_c}$ for $a=4,5,6,7$, and order $1$ for $a=8$.  Relative to this leading term, the terms $g_1$, $g_2$, $g_4$ and $g_5$ are order $1/N_c$ and the terms $g_3$, $g_6$ and $g_{7}$ are order $1/N_c^2$.  The coupling $g_A$ is normalized to the axial coupling of a constituent quark, so that $g_A(n \to p)=1.25=5g_A/3$ gives $g_A \sim 0.75$~\cite{georgi}. To leading order the axial pion octet couplings in the $E=0$ nucleon tower and $E=1$ exotic tower are equal; they differ at relative order $1/N_c$ due to $g_1$, $g_2$, $g_4$ and $g_5$.

The $\Delta E=-1$ axial currents are given by
\begin{eqnarray}
\frac 1 2 A^{ia}_-  &=&  \bar g_0 G_-^{ia} +  \bar g_1{{\{E, G_-^{ia}\}} \over N_c}  + \bar g_3 {J_q^i T^a_- \over N_c} + \bar g_4 {J_{\bar q}^i T^a_- \over N_c} \ \
\label{31}
\end{eqnarray}
to relative order $1/N_c$. The leading term $G^{ia}_-$ is order $\sqrt{N_c}$ for $a=1,2,3$, order $1$ for $a=4,5,6,7$, and order $1/\sqrt{N_c}$ for $a=8$, 
so these transition couplings are suppressed by $1/\sqrt{N_c}$ relative to the $\Delta E=0$ axial couplings, such as $g_{NN\pi}$, discussed above. The $1/N_c$ expansion of the $\Delta E= -1$ axial couplings relates the pion octet couplings between pentaquark and non-exotic baryons.  All of the decay couplings between the $E=1$ and $E=0$ baryons are given by a single unknown coupling $\bar g_0$ up to corrections of relative order $1/N_c$. A quantitative measure of the suppression of $\Delta E=-1$ couplings is given by the ratio $\bar g_0/g_A$.

\section{Mass Relations}\label{sec:masses}

A detailed discussion of the above equations including $SU(3)$ breaking will be given in  longer publications~\cite{JM3,Golbeck}. Here we limit ourselves to a discussion of the mass relations and widths in the $SU(3)$ flavor symmetry limit.
The Hamiltonian Eq.~(\ref{20}) yields, by taking linear combinations to eliminate the unknown coefficients $c_{0-5}$, the mass relations
\begin{eqnarray}
2 \left( \mathbf{\overline{10}}_{1/2}\right) + \vev{\mathbf{35}} &=& 3 \vev{\mathbf{27}} +   \mathcal{O}\left(1/N_c^3\right) ,\nn
\left(\mathbf{35}_{5/2}- \mathbf{35}_{3/2}\right) &=&  \frac 5 3  \left(\mathbf{27}_{3/2}- \mathbf{27}_{1/2}\right)    + \mathcal{O}\left(1/N_c^3\right),\nn
\vev{\mathbf{27}} - \mathbf{\overline{10}}_{1/2} &=& \frac 2 3 \left( \mathbf{10}_{3/2} - \mathbf{8}_{1/2} \right) + \mathcal{O}\left(1/N_c^2\right), \nn
\vev{\mathbf{35}} - \mathbf{\overline{10}}_{1/2} &=& 2 \left( \mathbf{10}_{3/2} - \mathbf{8}_{1/2} \right) + \mathcal{O}\left(1/N_c^2\right) ,
\label{results}
\end{eqnarray}
where
\begin{eqnarray}
\vev{\mathbf{27}} &\equiv& \frac 1 3 \left[ \left(\mathbf{27}_{1/2}\right) + 2 \left(\mathbf{27}_{3/2}\right) \right], \\
\vev{\mathbf{35}} &\equiv& \frac 1 5 \left[ 2 \left(\mathbf{35}_{3/2}\right) + 3 \left(\mathbf{35}_{5/2}\right) \right]
\end{eqnarray} 
denote spin-averaged masses. The fourth relation is not linearly independent. 
The first two relations also are satisfied by the $1/N_c^2$ terms in the Hamiltonian, and so are true to $1/N_c^3$.

Using only isospin $SU(2)$ flavor symmetry, one gets the corresponding relations for the $S=1$ states:
\begin{eqnarray}
2 \Theta^+  + \vev{\Theta_2} &=& 3 \vev{\Theta_1} +   \mathcal{O}\left(1/N_c^3\right) ,\nn
\left(\Theta_{2,j=5/2}-\Theta_{2,j=3/2}\right) &=&  \frac 5 3  \left(\Theta_{1,j=3/2}- \Theta_{1,j=1/2}\right)    + \mathcal{O}\left(1/N_c^3\right),\nn
\vev{\Theta_1} - \Theta^+ &=& \frac 2 3 \left( \Delta- N \right) + \mathcal{O}\left(1/N_c^2\right)=195~\hbox{MeV} + \mathcal{O}\left(1/N_c^2\right), \nn
\vev{\Theta_2} - \Theta^+ &=& 2 \left( \Delta -N \right) + \mathcal{O}\left(1/N_c^2\right)=586~\hbox{MeV} + \mathcal{O}\left(1/N_c^2\right) ,
\end{eqnarray}
where
\begin{eqnarray}
\vev{\Theta_1} &\equiv& \frac 1 3 \left( \Theta_{1, j=1/2} + 2 \Theta_{1, j=3/2} \right), \\
\vev{\Theta_2} &\equiv& \frac 1 5 \left( 2 \Theta_{2, j=3/2} + 3 \Theta_{2, j=5/2} \right),
\end{eqnarray} 
denote spin-averaged masses.
These relations do not rely on the assumption of approximate $SU(3)$ flavor symmetry, and they are identical to the mass relations obtained in Ref.~\cite{IKOR} in the Skyrme model.  The $1/N_c$ analysis given here is model-independent and gives the accuracy of each mass relations as a power of $1/N_c$.
The $1/N_c^3$ and $1/N_c^2$ mass relations are expected to hold at the 10~MeV and 30~MeV level since the leading order $N_c$ mass is approximately 1~GeV.
Using  $\Theta^+=1540$~MeV, these relations imply $\vev{\Theta_1}=1735$~MeV 
and $\vev{\Theta_2}=2126$~MeV, to an accuracy of $1/N_c^2$, or approximately $30$~MeV. One cannot determine the $\Theta^+$ mass because of the order $N_c^0 =1$ term linear in $E$ in the Hamiltonian which splits the exotic $E=1$ baryon masses from the normal $E=0$ baryon masses.\footnote{The decay widths $\Theta \to NK$ are of order $N_c^0$~\cite{Praszalowicz:2003tc,JM1} but do not destroy the above mass relations.  Decay couplings are constrained by the $1/N_c$ expansions Eq.~(\ref{30},\ref{31}), and yield decay widths which satisfy the same $1/N_c$ relations  as the masses. Mixing between exotic and ordinary baryons also does not destroy the hierarchy of baryon mass predictions for the ordinary baryons, including the successful Gell-Mann--Okubo relations. The mixing Hamiltonian Eq.~(\ref{31}) is constrained by the $1/N_c$ expansion, and produces mass shifts which satisfy the $1/N_c$ relations in Ref.~\cite{JenkinsLebed}.  In addition, it is known for ground-state baryons that radiative corrections such as chiral loops respect the $1/N_c$ relations~\cite{DJM1,loops}. This result should not be a surprise---the $1/N_c$ relations are \emph{symmetry} relations which are preserved by all interactions in theory.}

\section{Decay Widths}

The decay widths can be computed to order $1/N_c$ using Eqs.~(\ref{30},\ref{31}) for the pion octet couplings. The exotic and ground state baryons have the same parity, so the decays are in the odd partial waves, since the $\pi$ and $K$ are pseudoscalars.
The $j=1/2 \leftrightarrow j=1/2$ and $j=1/2 \leftrightarrow j=3/2$ transitions are only $p$-wave, by angular momentum conservation. All the decays are dominantly $p$-wave decays as  long as the momentum of the final particles is small. Transitions between normal baryons or between $E=1$ baryons are $\Delta E=0$ transitions, and depend on $g_A$. Transitions between the $E=1$ tower and the $E=0$ tower are $\Delta E=-1$ transitions, and depend on $\bar g_0$ (see Fig.~\ref{fig:1}). \FIGURE{
\psset{unit=0.75cm}
\begin{pspicture}(-2,0)(10,8)
\psline(0,0)(1,0)
\rput(-1,0){$N$}
\psline(0,2)(1,2)
\rput(-1,2){$\Delta$}
\psline(6,3)(7,3)
\rput(8,3){$\Theta^+$}
\psline(6,5)(7,5)
\rput(8,5){$\Theta_1$}
\psline(6,7)(7,7)
\rput(8,7){$\Theta_2$}
\psline[linecolor=blue]{->}(6.5,3)(0.5,0)
\rput(4,1){$K$}
\psline[linecolor=blue]{->}(6.5,5)(0.5,0)
\psline[linecolor=blue]{->}(6.5,5)(0.5,2)
\psline[linecolor=blue]{->}(6.5,7)(0.5,2)
\psline[linecolor=red]{->}(0.5,2)(0.5,0)
\psline[linecolor=red]{->}(6.5,5)(6.5,3)
\psline[linecolor=red]{->}(6.5,7)(6.5,5)
\rput(1,1){$\pi$}
\rput(7,4){$\pi$}
\rput(7,6){$\pi$}
\end{pspicture}
\caption{Transitions between the pentaquark and normal baryon towers. The transitions within a tower (vertical red lines) depend on $g_A$, and those between towers (diagonal blue lines) depend on $\bar g_0$. The $\Theta^+$ is naturally much narrower than the excited pentaquarks, since it is the lightest pentaquark and cannot decay within the pentaquark tower. \label{fig:1}}
}

The $\Delta E=0$ transitions can be normalized to $\Delta \to N\pi$, and the $\Delta
E = -1$ transitions can be normalized to $\Theta \to N K$.
The decay widths all have a $p^3$ phase space dependence, where $p$ is the momentum of the emitted pseudoscalar meson. The ratios of decay widths correcting for the $p^3$ phase space dependence (i.e.\ the ratios of the transition amplitudes squared) are:
\begin{eqnarray}
 {p^{-3} \Gamma( \Theta_{1,j=1/2} \to NK ) \over p^{-3}\Gamma(\Theta \to N K)}=\frac 2 9,\nn
{p^{-3}\Gamma( \Theta_{1,j=3/2} \to NK ) \over p^{-3}\Gamma(\Theta \to N K)} =\frac 8 9, \nn
{p^{-3}\Gamma( \Theta_{1,j=1/2} \to \Delta K ) \over p^{-3}\Gamma(\Theta \to N K)}= \frac 4 9,\nn
{p^{-3}\Gamma( \Theta_{1,j=3/2} \to \Delta K ) \over p^{-3}\Gamma(\Theta \to N K)} = \frac 5 {18},  \nn
{p^{-3}\Gamma( \Theta_{2,j=3/2} \to \Delta K ) \over p^{-3} \Gamma(\Theta \to N K)} =\frac12,\nn
{p^{-3}\Gamma( \Theta_{2,j=5/2} \to \Delta K ) \over p^{-3} \Gamma(\Theta \to N K) }=\frac 43,\nn
{p^{-3} \Gamma( \Theta_{1,j=1/2} \to \Theta \pi ) \over p^{-3}\Gamma(\Delta \to N \pi) }= \frac 43,\nn
{p^{-3} \Gamma( \Theta_{1,j=3/2} \to \Theta \pi ) \over p^{-3} \Gamma(\Delta \to N \pi) } =\frac43, \nn
{p^{-3} \Gamma( \Theta_{2,j=3/2} \to \Theta_{1,j=1/2} \pi ) \over p^{-3} \Gamma(\Delta \to N \pi) } = \frac 5 4,\nn
{p^{-3} \Gamma( \Theta_{2,j=3/2} \to \Theta_{1,j=3/2} \pi ) \over p^{-3} \Gamma(\Delta \to N \pi) } = \frac 1 4 , \nn
{p^{-3} \Gamma( \Theta_{2,j=5/2} \to \Theta_{1,j=3/2} \pi ) \over p^{-3} \Gamma(\Delta \to N \pi) } = \frac 32 .
\label{pqwidths}
\end{eqnarray}
The pion decay ratios have corrections of order $1/N_c$, since they involve transitions in two different towers. The corrections are of order $1/N_c^2$ if one takes ratios of pion decays between different states in the pentaquark tower, e.g.\  $p^{-3}  \Gamma( \Theta_{1,j=1/2} \to \Theta \pi )/ p^{-3} \Gamma( \Theta_{1,j=3/2} \to \Theta \pi )=1 + \mathcal{O}(1/N_c^2)$, by taking the ratio of two pion entries in Eq.~(\ref{pqwidths}).
The kaon decay ratios have corrections of order $1/N_c$. 

The $\Delta$ width is~\cite{hungary}
\begin{eqnarray}
\Gamma(\Delta \to N \pi) &=& {g_A^2 p^3 \over 3  \pi f^2_\pi},
\end{eqnarray}
which normalizes all of the $\Delta E=0$ decays. The $\Theta$ width is
\begin{equation}
\Gamma(\Theta^+)=\Gamma(\Theta^+\to N K) = {\bar g_0^2 p^3\over 2 \pi f_K^2}  = 10~\text{MeV} \left({\bar g_0 \over 0.2}\right)^2,
\end{equation}
which shows that $\bar g_0/g_A$ must be at least a factor of four smaller than $g_A$ for a $\Theta^+$ width below 10~MeV. The coupling between different towers $\bar g_0$ is expected to be smaller than the coupling within a given tower $g_A$, $\bar g_0 < g_A$. An example of such a suppression is the comparison of the coupling between the excited nucleon and ground state nucleon towers: $g(N^*N\pi)^2/g(NN\pi)^2 \sim 1/10$~\cite{hungary}.

The $j=3/2$ and $5/2$ isotensors $\Theta_{2}$ and the $j=1/2$ and $3/2$ isotriplets $\Theta_{1}$ can decay via $\Delta E=0$ pion emission to lower states in the $E=1$ tower; the $\Theta$, being the lightest $E=1$ state with $S=1$, can only decay via a $\Delta E=-1$ transition, as show in Fig.~\ref{fig:1}. Transitions within a tower are controlled by $g_A$ and between towers by $\bar g_0$, so that naturally $\Gamma(\Theta) \ll \Gamma(\Theta_1),\Gamma(\Theta_2)$. Using 1735~MeV for both the $\Theta_1$ states, and 2126~MeV for both the $\Theta_2$ states gives $\Gamma(\Theta_{1,j=1/2}) \agt 30$~MeV + $1.2 \Gamma(\Theta)$, $\Gamma(\Theta_{1,j=3/2}) \agt 30$~MeV + $4.9 \Gamma(\Theta)$, $\Gamma(\Theta_{2,j=3/2})\agt 560$~MeV + $5.1 \Gamma(\Theta)$ and $\Gamma(\Theta_{2,j=5/2})\agt 560$~MeV + $13.6 \Gamma(\Theta)$.\footnote{A note of caution: The $\Theta_{1,j=3/2}-\Theta_{1,j=1/2}$ mass splitting is not included in the numerical values quoted. The $p^3$ phase-space factor depends strongly on the masses. The results given are inequalities because there may be other channels (such as $K$ decays between pentaquarks) which can exist if the mass splittings allow.}  
The isotensor $\mathbf{35}$ states are seen to be extremely broad, and will not be distinguishable from the continuum. 

The widths of the isotriplet flavor-$\mathbf{27}$ states depend on the value of the hyperfine splitting between the $j=1/2$ and $j=3/2$ states. Since the hyperfine splitting does not change the average mass, one isotriplet state will have higher mass (and width), and the other lower mass (and width) than the results quoted using the average mass. In Figs.~(\ref{fig:width},\ref{fig:br}) we have plotted the decay widths and $NK$ branching ratio of the states $\Theta_{1,j=1/2}$ and $\Theta_{1,j=3/2}$ as a function of their mass.
\DOUBLEFIGURE{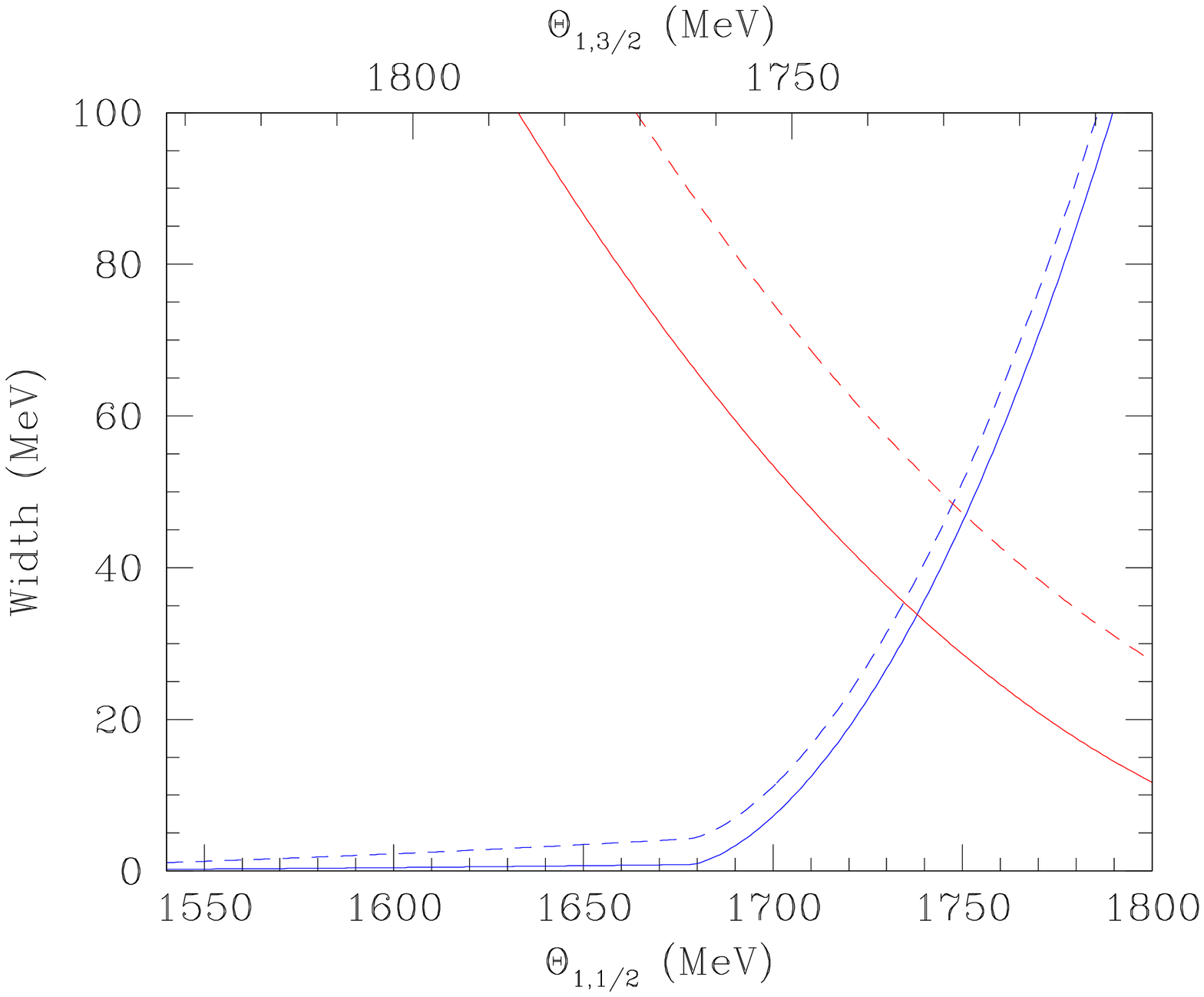,width=7cm}{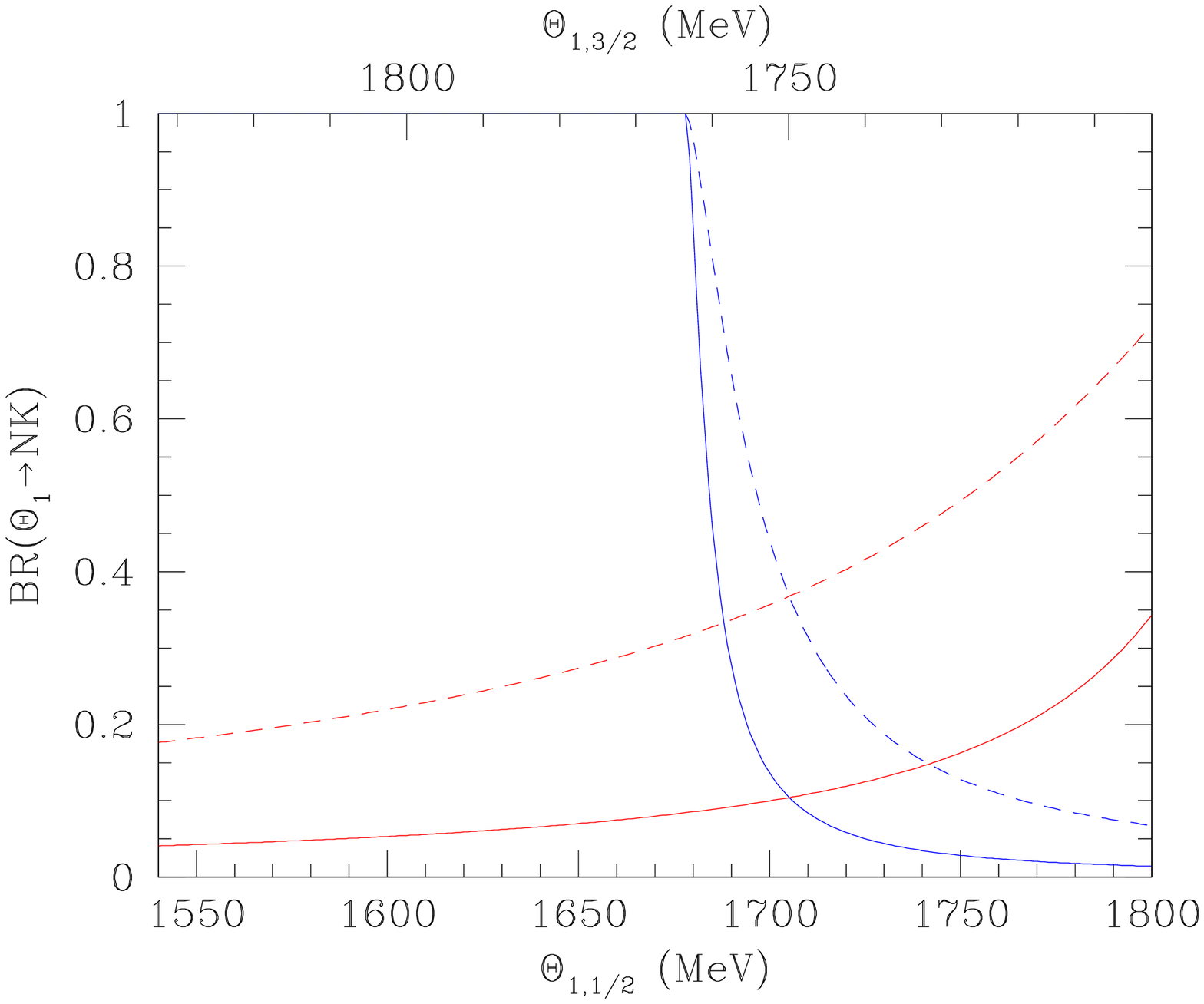,width=7cm}
{Width of the $\Theta_1$ states as a function of the $\Theta_1$ mass. The blue curves are for $j=1/2$, and the red curves for $j=3/2$.
The solid and dashed curves are for a $\Theta$ width of 1 and 5~MeV, respectively. \label{fig:width}}
{Branching ratio of the  $\Theta_1$ states into $NK$ as a function of the $\Theta_1$ mass.  The blue curves are for $j=1/2$, and the red curves for $j=3/2$.
The solid and dashed curves are for a $\Theta$ width of 1 and 5~MeV, respectively. \label{fig:br}}
The mass of the $j=1/2$ state is given by the lower axis, and the mass of the $j=3/2$ state by the upper axis. Note that the two masses are correlated---
the spin-averaged mass of the two states has been set to $1735$~MeV. The variation with mass is entirely due to the $p^3$ phase space dependence of the widths. The rapid rise in the $j=1/2$ width (and fall in its $NK$ branching ratio) around 1700~MeV is due to onset of the $\Theta \pi$ decay channel. The $\Theta_1$ states are wide as long as they are above $\Theta \pi$ threshold.

\section{Heavy Pentaquarks}

One can have $qqqq\bar Q$ exotics containing heavy quarks, which are analogs of the the $\Theta$ and related states. We will study the masses and decays of these heavy pentaquarks in this section.  Note that $qqqQ\bar q$ exotics~\cite{boundstate}, as well as exotics containing more than one heavy quark, also could exist, but are not discussed here.

\subsection{Masses}

One can derive mass relations for the lowest-lying $qqqq\bar Q$ baryons ($Q=c,b$), which consist of the $\mathbf{\overline{6}}_{1/2}$, $\mathbf{15}_{1/2}$, $\mathbf{15}_{3/2}$, 
$\mathbf{15^\prime}_{3/2}$ and $\mathbf{15^\prime}_{5/2}$ multiplets. The mass relations in the $SU(3)$ flavor symmetry limit are
\begin{eqnarray}
2 \left( \mathbf{\overline{6}}_{1/2}\right) +\vev{\mathbf{15^\prime}} &=& 3 \vev{\mathbf{15}}+
 \mathcal{O}\left(1/N_c^3\right), \nn
\left(\mathbf{15^\prime}_{5/2}- \mathbf{15^\prime}_{3/2}\right) &=&  \frac 5 3 \left(\mathbf{15}_{3/2}- \mathbf{15}_{1/2}\right) + \mathcal{O}\left(1/(m_Q N_c^3)\right) ,\nn
\vev{\mathbf{15}} - \mathbf{\overline{6}}_{1/2} &=& \frac 2 3 (\mathbf{10}_{3/2} - \mathbf{8}_{1/2})   + \mathcal{O}\left(1/N_c^2\right), \nn
\vev{\mathbf{15^\prime}} - \mathbf{\overline{6}}_{1/2} &=&  2  (\mathbf{10}_{3/2} - \mathbf{8}_{1/2}) + \mathcal{O}\left(1/N_c^2\right) , 
\end{eqnarray}
where
\begin{eqnarray}
\vev{\mathbf{15}} &=& \frac 1 3 \left[ \left(\mathbf{15}_{1/2}\right) + 2 \left(\mathbf{15}_{3/2}\right) \right], \\
\vev{\mathbf{15^\prime}} &=& \frac 1 5 \left[ 2 \left(\mathbf{15^\prime}_{3/2}\right)  + 3 \left(\mathbf{15^\prime}_{5/2}\right) \right],
\end{eqnarray} 
denote spin-averaged masses and
the $\mathbf{6}_{1/2}$ and $\mathbf{6}_{3/2}$ are the non-exotic $qqQ$ baryon multiplets containing the $\Sigma_{c,b}$ and $\Sigma_{c,b}^*$. 
The $\left(\mathbf{15}_{3/2}- \mathbf{15}_{1/2}\right)$ and $\left(\mathbf{15^\prime}_{5/2}- \mathbf{15^\prime}_{3/2}\right)$ mass splittings violate heavy quark spin-symmetry and are proportional to $1/m_Q$. The above mass relations are identical to those for the light pentaquarks, since the Hamiltonian has the same structure as Eq.~(\ref{20}). The only difference is that the hyperfine splittings, which violate heavy quark spin symmetry, are suppressed by $1/m_Q$.

For two light flavors, the $qqqq\bar Q$ baryon exotics consist of the isosinglet  
$\Theta_{\bar Q}$ in the $\mathbf{\overline{6}}_{1/2}$, the isotriplets $\Theta_{1,\bar Q, j=1/2}$ and $\Theta_{1,\bar Q, j=3/2}$ in the $\mathbf{15}_{1/2}$ and $\mathbf{15}_{3/2}$, and the isotensors $\Theta_{2,\bar Q, j=3/2}$ and $\Theta_{2,\bar Q, j=3/2}$ in the $\mathbf{15^\prime}_{3/2}$ and $\mathbf{15^\prime}_{5/2}$.
Using only isospin $SU(2)$ flavor symmetry, one obtains the mass relations
\begin{eqnarray}
2 \left(\Theta_{\bar Q} \right) +\vev{\Theta_{2,\bar Q}} &=& 3 \vev{\Theta_{1,\bar Q}}+
 \mathcal{O}\left(1/N_c^3\right), \nn
\left(\Theta_{2,\bar Q,j=5/2}- \Theta_{2,\bar Q,j=3/2}\right) &=&  \frac 5 3 \left(\Theta_{1,\bar Q,j=3/2}- \Theta_{1,\bar Q,j=1/2}\right) + \mathcal{O}\left(1/(m_Q N_c^3)\right) ,\nn
\vev{\Theta_{1,\bar Q}} - \Theta_{\bar Q} &=& \frac 2 3 (\Delta -N)   + \mathcal{O}\left(1/N_c^2\right), \nn
\vev{\Theta_{2,\bar Q}} - \Theta_{\bar Q} &=&  2  (\Delta-N) + \mathcal{O}\left(1/N_c^2\right), 
\end{eqnarray}
where
\begin{eqnarray}
\vev{\Theta_{1,\bar Q}} &\equiv& \frac 1 3 \left( \Theta_{1, \bar Q, j=1/2} + 2 \Theta_{1, \bar Q, j=3/2} \right), \\
\vev{\Theta_{2,\bar Q}} &\equiv& \frac 1 5 \left( 2 \Theta_{2, \bar Q, j=3/2} + 3 \Theta_{2,
\bar Q, j=5/2} \right),
\end{eqnarray}
denote spin-averaged masses.
These mass relations imply
\begin{eqnarray}
\Theta_{1,\bar Q, j=1/2} &=& \Theta_{\bar Q} + \frac 2 3 \left(\Delta - N\right)-\frac 2 3  \delta {\Theta_{1,\bar Q}} +  \mathcal{O}\left(1/N_c^2\right) ,\nn
\Theta_{1,\bar Q, j=3/2} &=& \Theta_{\bar Q} + \frac 2 3 \left(\Delta - N\right)+\frac 1 3 \delta {\Theta_{1,\bar Q}} +  \mathcal{O}\left(1/N_c^2\right), \nn
\Theta_{2,\bar Q, j=3/2} &=& \Theta_{\bar Q} + 2 \left(\Delta - N\right)-\delta {\Theta_{1,\bar Q}}  +  \mathcal{O}\left(1/N_c^2\right), \nn
\Theta_{2,\bar Q, j=5/2} &=& \Theta_{\bar Q} + 2 \left(\Delta - N\right)+\frac 23 \delta {\Theta_{1,\bar Q}}  +  \mathcal{O}\left(1/N_c^2\right) .
\end{eqnarray}
where $\delta {\Theta_{1,\bar Q}}=\left(\Theta_{1,\bar Q,j=3/2}- \Theta_{1,\bar Q,j=1/2}\right)$ is the heavy quark spin-symmetry violating hyperfine splitting proportional to $(J_{\bar Q} \cdot J_q)/ N_c m_Q$~\cite{heavybaryons}.
Substituting in the numerical value for the $\left(\Delta-N\right)$ mass difference gives
\begin{eqnarray}
\Theta_{1,\bar c, j=1/2} &=& \Theta_{\bar c} +
195~\hbox{MeV} -\frac 23 \delta {\Theta_{1,\bar c}}+  \mathcal{O}\left(1/N_c^2\right), \nn
\Theta_{1,\bar c, j=3/2} &=& \Theta_{\bar c} +
195~\hbox{MeV} + \frac 1 3 \delta {\Theta_{1,\bar c}}  + \mathcal{O}\left(1/N_c^2\right), \nn
\Theta_{2,\bar c, j=3/2} &=& \Theta_{\bar c} +
586~\hbox{MeV} -\delta {\Theta_{1,\bar c}} +  \mathcal{O}\left(1/N_c^2\right) ,\nn
\Theta_{2,\bar c, j=5/2} &=& \Theta_{\bar c} +
586~\hbox{MeV} + \frac 2 3 \delta {\Theta_{1,\bar c}}+  \mathcal{O}\left(1/N_c^2\right),
\end{eqnarray}
for the isotriplet and isotensor charmed exotics. The $(J_Q \cdot J_q)/N_c m_Q$ hyperfine splitting $\Sigma_c^*-\Sigma_c=65.6~\hbox{MeV}$ also violates heavy quark spin-symmetry, and gives an estimate of the size of $\delta {\Theta_{1,\bar c}}$. The $1/N_c$ expansion does not allow one to predict the mass of the $\Theta_{\bar c}$ to great accuracy; a relation such as $\Theta_{\bar c} - \Lambda_c=\Theta - \Lambda$ has an order $1/N_c$ correction, and gives a $\Theta_{\bar c}$ mass of $2700$~MeV with corrections of order $1/N_c$. An order $1/N_c$ mass term is typically of order 100~MeV.  The recently reported charm pentaquark at 3099~MeV~\cite{h1} is 400~MeV above the estimated $\Theta_{\bar c}$ mass, and so is
unlikely to be the ground state pentaquark. Other estimates of the ground state charm pentaquark mass~\cite{Jaffe,sww} also give a value much smaller than 3099~MeV.
If we identify the resonance at 3099~MeV with one of the excited $\Theta_{1,\bar c}$ pentaquarks, then the mass of the ground state charm pentaquark $\Theta_{\bar c}$ is more accurately given by $2904{+(2/3)\delta {\Theta_{1,\bar c}} \atop -(1/3)\delta {\Theta_{1,\bar c}}}$~MeV, depending on whether the observed $\Theta_{1,\bar c}$ pentaquark has $j=1/2$ or $3/2$.  This mass result has a $1/N_c^2$ correction, which is estimated to be 30~MeV.

\subsection{Widths}

The decay widths of the heavy pentaquarks can be studied using the $1/N_c$ expansion and heavy quark symmetry.  Heavy pentaquarks can decay within the heavy pentaquark tower by pion emission and to normal baryons plus heavy pseudoscalar and vector mesons $P_{\bar Q}$ and $P_{\bar Q}^*$.

The transition amplitudes for pion decays of heavy pentaquarks are equal to those of the corresponding light pentaquarks,  e.g.\
$p^{-3} \Gamma( \Theta_{1,\bar Q,j=1/2} \to \Theta_{\bar Q} \pi)  = p^{-3} \Gamma( \Theta_{1,j=1/2} \to \Theta \pi )$. One can also use Eq.~(\ref{pqwidths}) to write these 
as relations between the heavy pentaquark and $\Delta$ widths, e.g.\ $p^{-3} \Gamma( \Theta_{1,\bar Q,j=1/2} \to \Theta_{\bar Q} \pi) = (4/3) p^{-3}\Gamma(\Delta \to N \pi) $. One expects the former version to have smaller $1/N_c$ corrections.

The decay widths $\Gamma(\Theta_{1, \bar c, j=1/2} \to \Theta_{\bar c} \pi)$ and $\Gamma(\Theta_{1, \bar c, j=3/2} \to \Theta_{\bar c} \pi)$ are 
equal to $ 0.26\, \Gamma(\Delta \to N \pi) \sim 30$~MeV, using 195~MeV for the mass difference between the $\Theta_{1\bar c}$ and $\Theta_{\bar c}$ states. The two decay rates are equal by heavy quark symmetry, since the decays involve transitions among the light degrees of freedom, and do not affect the heavy quark. The hyperfine splitting $\delta {\Theta_{1,\bar c}}$ violates heavy quark spin symmetry and changes the widths, since it changes the $p^3$ phase-space dependence. The hyperfine interaction is much smaller for the heavier pentaquarks, and so is less important than in the light pentaquark sector. The $\Theta_{\bar c, 2} \to \Theta_{\bar c, 1} \pi$ widths are extremely broad, and the states will not be observable. 

The decay widths of heavy pentaquarks $\Theta_{\bar Q}$ to nucleon plus pseudoscalar and vector mesons, $N+ P_{\bar Q}$ and $N+ P_{\bar Q}^*$, can be related using heavy quark spin-symmetry and the $1/N_c$ expansion.
We first determine the relations which follow from heavy quark symmetry \emph{alone}, and then combine them with the $1/N_c$ results.

Heavy quark spin symmetry relations are derived between the various decay amplitudes. Whether these are allowed decays depends on the masses of the states.
The charm pentaquark reported by H1 can decay into both  $N+\bar D$ and $N+\bar D^*$. If it is an excited pentaquark, then some of these decay channels may be kinematically forbidden for the $\Theta_{\bar c}$, which is lighter. It is also possible that the $\Theta_{\bar c}$ has no strong decays, and only decays via the weak interactions~\cite{Jaffe,sww}.
The $\Theta_{2,\bar Q}$ states are likely to be too broad to be seen, so we derive 
results only for the $\Theta_{\bar Q}$ and $\Theta_{1,\bar Q}$. 

The $\Theta_{\bar Q}$ has $J=1/2$ and $J_\ell=0$, where $J_\ell$ is the spin of the light degrees of freedom. It is represented by the spinor field $u_\Theta$. The
$\Theta_{1,\bar Q}$ states have $J=1/2,3/2$ and $J_\ell=1$, and are represented by the spinor field~\cite{Falk}
\begin{eqnarray}
u^\mu_{\Theta,1} &=& {1\over \sqrt{3}} \left( \gamma^\mu + v^\mu \right) \gamma_5 u_{\Theta,1,j=1/2}
+ u^\mu_{\Theta,1,j=3/2} .
\label{50}
\end{eqnarray}
The pseudoscalar and vector mesons are represented by the $H^{(\bar Q)}$ field~\cite{book}. To simplify the notation, in any given decay, the initial pentaquark is denoted by $\Theta$, and the final meson by $H$.

The decay amplitudes are computed in the rest frame of the decaying pentaquark, so that $v=(1,0,0,0)$, and the momentum of the initial state is $p_\Theta=m_\Theta v$. The decay products are non-relativistic, so the velocity of the heavy quark in the heavy meson is also chosen to be $v$. The momentum of the heavy meson is given by $p_H = m_H v + k$, where $k$ is the residual momentum. The momentum of the nucleon is $p_N$. Momentum conservation implies that $m_\Theta v = m_H v + k + p_N$, so we have two independent four-vectors, which can be chosen to be $v$ and $p_N$.

The most general invariant amplitude for $\Theta_{\bar Q} \to N + H_{\bar Q}$ consistent with heavy quark symmetry is $\bar u_N \, \Gamma \, \bar H^{(\bar Q)}\, u_\Theta$
where $\Gamma$ is an arbitrary bispinor matrix which can be constructed out of $v$, $p_N$ and products of $\gamma$ matrices, and $u_N$ is the nucleon spinor. Using $\xslash p_N u_N=m_N u_N$ and $\xslash v H^{(\bar Q)} = H^{(\bar Q)}$, one can reduce all the terms to a single invariant with the correct parity and time-reversal properties,
\begin{eqnarray}
f_0\ \bar u_N \,  \, \bar H^{(\bar Q)}\, u_\Theta
\label{51}
\end{eqnarray}
where $f_0$ is an unknown constant. That there is only one invariant follows by counting the ways in which the spins of the light degrees of freedom in the final state can be combined to form $J_\ell=0$.

Similarly, for the $\Theta_{1,\bar Q}$ states, one can show that there are only two invariants, which are chosen to be
\begin{eqnarray}
\left[ f_1\ \bar u_N \gamma_\mu \gamma_5 \bar H^{(\bar Q)} + f_2 \ \bar u_N p_{N,\mu} \gamma_5\bar H^{(\bar Q)} \right]  u^\mu_\Theta
\label{53}
\end{eqnarray}
where $ u^\mu_\Theta$ is given in Eq.~(\ref{50}). A third possible invariant involving the $\epsilon$ symbol exists, but is not linearly independent,
\begin{eqnarray}
(m_N + p_N \cdot v) \bar u_N \gamma_\mu \gamma_5 \bar H^{(\bar Q)}u^\mu_\Theta
&=&
i r \epsilon_{ \alpha \beta \nu \mu} v^\alpha p_{N,\beta} \bar u_N \gamma^\nu H^{(\bar Q)} u^\mu_\Theta +  \ \bar u_N p_{N,\mu} \gamma_5\bar H^{(\bar Q)}  u^\mu_\Theta .\nn
\label{52}
\end{eqnarray}

The six possible decays of the $\Theta_{\bar Q}$ and $\Theta_{1,\bar Q}$ into the 
$P_{\bar Q}$ and $P_{\bar Q}^*$ mesons are given in terms of three coupling constants $f_{0,1,2}$,
\begin{eqnarray}
\Gamma \left(\Theta_{\bar Q} \to P_{\bar Q} + N \right)  &=&  {\abs{\mathbf{p}_N} \left(E_N - m_N \right)  E_D \over 2 \pi m_\Theta } f_0^2,    \nn
\Gamma \left(\Theta_{\bar Q} \to P_{\bar Q}^* + N \right)  &=&   {\abs{\mathbf{p}_N} \left(E_N - m_N \right)  E_D \over 2 \pi m_\Theta } 3 f_0^2,  \nn
\Gamma \left(\Theta_{1,\bar Q,j=1/2} \to P_{\bar Q} + N \right)  &=&   {\abs{\mathbf{p}_N} \left(E_N - m_N \right)  E_D \over 2 \pi m_\Theta }    \frac 1 3 \left[3 f_1  +  \tilde f_2 \right]^2   ,\nn
\Gamma \left(\Theta_{1,\bar Q,j=3/2} \to P_{\bar Q} + N \right)  &=&  {\abs{\mathbf{p}_N} \left(E_N - m_N \right)  E_D \over 2 \pi m_\Theta } \frac 1 3 \tilde f_2^2 , \nn
\Gamma \left(\Theta_{1,\bar Q,j=1/2} \to P_{\bar Q}^* + N \right)  &=&   {\abs{\mathbf{p}_N} \left(E_N - m_N \right)  E_D \over 2 \pi m_\Theta } \Bigl[  f_1^2  + \frac 2 3 f_1 \tilde f_2  + \tilde f_2^2 \Bigr]  ,\nn
\Gamma \left(\Theta_{1,\bar Q,j=3/2} \to P_{\bar Q}^* + N \right)  &=&  {\abs{\mathbf{p}_N} \left(E_N - m_N \right)  E_D \over 2 \pi m_\Theta } \Bigl[4 f_1^2+ \frac {8} 3 f_1 \tilde f_2 + \tilde f_2^2  \Bigr] ,
\label{hqw}
\end{eqnarray}
where $\tilde f_2 = (m_N+E_N) f_2$, and we have normalized the amplitudes in Eqs.~(\ref{50},\ref{51}) so that the isospin Clebsch-Gordan factors are unity for each decay. There are three relations among the decay rates, which follow from heavy quark spin symmetry (and do not use the $1/N_c$ expansion),
\begin{eqnarray}
p^{-3}\Gamma \left(\Theta_{\bar Q} \to P_{\bar Q}^* + N \right)  &=& 3 p^{-3} \Gamma \left(\Theta_{\bar Q} \to P_{\bar Q} + N \right) ,\nn
p^{-3} \Gamma \left(\Theta_{1,\bar Q,j=1/2} \to P_{\bar Q}^* + N \right)  &=&
\frac 1 3 p^{-3} \Gamma \left(\Theta_{1,\bar Q,j=1/2} \to P_{\bar Q} + N \right) +
\frac 8 3 p^{-3} \Gamma \left(\Theta_{1,\bar Q,j=3/2} \to P_{\bar Q} + N \right) ,\nn
p^{-3}\Gamma \left(\Theta_{1,\bar Q,j=3/2} \to P_{\bar Q}^* + N \right) &=&
\frac 4 3 p^{-3}\Gamma \left(\Theta_{1,\bar Q,j=1/2} \to P_{\bar Q} + N \right) +
\frac 5 3 p^{-3}\Gamma \left(\Theta_{1,\bar Q,j=3/2} \to P_{\bar Q} + N \right),\nn
\label{hqwidth}
\end{eqnarray}
where we have used $E_N - m_N \approx p_N^2/(2m_N)$, since the nucleon is non-relativistic. In these relations $p$ is the three-momentum of the final state particle for the corresponding decay. These relations are violated by $1/m_Q$ corrections.

In addition to the heavy quark relations Eq.~(\ref{hqwidth}), there are relations which follow from isospin symmetry. The $\Theta_Q$ is isospin zero, and so
\begin{eqnarray}
\Gamma \left(\Theta_{\bar Q} \to H(d \bar Q) + p \right)=\Gamma \left(\Theta_{\bar Q} \to H(u \bar Q) + n \right) = \frac 1 2 \Gamma \left(\Theta_{\bar Q} \to H_{\bar Q} + N \right)
\end{eqnarray}
where $H(q \bar Q)$ represents either the pseudoscalar or the vector meson with quark content $q \bar Q$, and the right hand side is the total width into the corresponding channel. The $\Theta_{1,\bar Q}$ has isospin one, so the $I_3=1$ pentaquark decays entirely into $H(u \bar Q)+p$, the $I_3=-1$ pentaquark decays into $H(d \bar Q)+n$ and the $I_3=0$ pentaquark decays 50\% of the time into $H(d \bar Q) + p$ and 50\% of the time into $H(u \bar Q) + n$.

The $1/N_c$ relations for the heavy pentaquark decays into pseudoscalar mesons are identical to the relations in Eq.~(\ref{pqwidths}), with the replacement $K \to P_{\bar Q}$. 
Combining the heavy quark predictions Eq.~(\ref{hqwidth}) with the large $N_c$ relations gives the decay width ratios:
\begin{eqnarray}
&& p^{-3} \Gamma \left(\Theta_{\bar Q} \to P_{\bar Q} + N \right) 
: p^{-3} \Gamma \left(\Theta_{\bar Q} \to P_{\bar Q}^* + N \right)  :
p^{-3} \Gamma \left(\Theta_{1,\bar Q,j=1/2} \to P_{\bar Q} + N \right)  \nn
&& : p^{-3} \Gamma \left(\Theta_{1,\bar Q,j=3/2} \to P_{\bar Q} + N \right)  
: p^{-3} \Gamma \left(\Theta_{1,\bar Q,j=1/2} \to P_{\bar Q}^* + N \right) \nn
&& : p^{-3}\Gamma \left(\Theta_{1,\bar Q,j=3/2} \to P_{\bar Q}^* + N \right)  \nn
&& = 1 : 3 : \frac 2 9 : \frac 8 9 : \frac{22} 9 : \frac {16} 9.
\label{61}
\end{eqnarray}
These relations are violated by $1/N_c^2$ and $1/m_c$ corrections.

In the large-$N_c$ limit, the method of Ref.~\cite{heavybaryons} gives the heavy pentaquark-baryon-heavy meson couplings in terms of a single coupling.  The large-$N_c$ spin-flavor symmetry implies that $\tilde f_2 = - 2 f_1$ and $f_1 = \sqrt{2/3} f_0$, which gives another way to derive Eq.~(\ref{61}).

One can similarly work out the consequences of heavy quark symmetry for the decays of negative parity heavy pentaquark states. The decay amplitudes are given by replacing Eq.~(\ref{51})  by $f_0\ \bar u_N \,i \gamma_5 \, \bar H^{(\bar Q)}\, u_\Theta$ and dropping the $\gamma_5$ in both terms in Eq.~(\ref{53}). The decay widths are given by
the replacement $m_N \to -m_N$ in Eq.~(\ref{hqw}) and in the definition of $\tilde f_2$. The phase space is now proportional to $\abs{\mathbf{p}}$, since the decays are predominantly $s$-wave. The relations Eq.~(\ref{hqwidth}) are replaced by
\begin{eqnarray}
p^{-1}\Gamma \left( \tilde \Theta_{\bar Q} \to P_{\bar Q}^* + N \right)  &=& 3 p^{-1} \Gamma \left(\tilde \Theta_{\bar Q} \to P_{\bar Q} + N \right) ,\nn
p^{-1} \Gamma \left(\tilde \Theta_{1,\bar Q,j=1/2} \to P_{\bar Q}^* + N \right)  &=&
\frac 1 3 p^{-1} \Gamma \left(\tilde \Theta_{1,\bar Q,j=1/2} \to P_{\bar Q} + N \right),\nn
p^{-1}\Gamma \left(\tilde \Theta_{1,\bar Q,j=3/2} \to P_{\bar Q}^* + N \right) &=&
\frac 4 3 p^{-1}\Gamma \left(\tilde \Theta_{1,\bar Q,j=1/2} \to P_{\bar Q} + N \right) ,\nn
 p^{-1}\Gamma \left(\tilde \Theta_{1,\bar Q,j=3/2} \to P_{\bar Q}+N\right) &=& \mathcal{O}\left( \abs{\mathbf{p}_N}^4 \right),
\label{71}
\end{eqnarray}
since $\tilde f_2 \approx \abs{\mathbf{p}_N}^2 f_2/(2M_N) $ is small. We have denoted the negative parity states by $ \tilde \Theta_{\bar Q}$ and $ \tilde \Theta_{1,\bar Q}$, which have $J_\ell=0$ and $J_\ell=1$, respectively. The last decay rate is $\mathcal{O}\left( \abs{\mathbf{p}_N}^4 \right)$, since it is a $d$-wave decay.

\acknowledgments

AM would like to thank Frank Wuerthwein for discussions on heavy pentaquarks. This work was supported in part by the Department of Energy under Grant 
DE-FG03-97ER40546.

\end{document}